\documentclass[aps,prxquantum,twocolumn,reprint,superscriptaddress]{revtex4-2}
\pdfoutput=1
\usepackage[utf8]{inputenc}
\usepackage[english]{babel}
\usepackage{csquotes}
\usepackage[stable]{footmisc}
\usepackage{hyperref}
\usepackage{amsmath, amsfonts, amssymb, dsfont}
\usepackage{graphicx}
\usepackage{subcaption}
\usepackage{adjustbox}
\usepackage{wrapfig}
\usepackage{booktabs}
\usepackage{nicefrac}
\usepackage{microtype}
\usepackage{xcolor}
\usepackage{lipsum}

\renewcommand{\vec}[1]{\mathbf{#1}}

\begin{document}

\title{Modeling the Quantum Photon Statistics in Hybrid Light-Matter Integrated Circuits}

\author{Mathias Van Regemortel}
\email{mathias.van-regemortel@hpe.com}
\affiliation{Large-Scale Integrated Photonics Lab, Hewlett Packard Labs, HPE Belgium, Diegem, Belgium}

\author{Vincenzo Ardizzone}
\email{vincenzo.ardizzone@cnr.it}
\affiliation{Institute of Nanotechnology, National Research Council, campus Ecotekne, Lecce 73100, Italy}

\author{Eugenio Maggiolini}
\affiliation{Institute of Semiconductor and Solid State Physics, Johannes Kepler University, Altenberger Stra\ss{}e 69, Linz 4040, Austria}

\author{Armando Rastelli}
\affiliation{Institute of Semiconductor and Solid State Physics, Johannes Kepler University, Altenberger Stra\ss{}e 69, Linz 4040, Austria}

\author{Daniele Sanvitto}
\affiliation{Institute of Nanotechnology, National Research Council, campus Ecotekne, Lecce 73100, Italy}

\author{Thomas Van Vaerenbergh}
\affiliation{Large-Scale Integrated Photonics Lab, Hewlett Packard Labs, HPE Belgium, Diegem, Belgium}

\begin{abstract}
Strong light-matter coupling between a guided electromagnetic mode and an excitonic semiconductor transition gives rise to exciton-polaritons with optical nonlinearities far exceeding those of conventional photonic platforms. Utilizing these nonlinearities in the few-particle regime, where quantum signatures such as photon antibunching, sub-Poissonian statistics and non-trivial inter-mode correlations become accessible, is a central goal of integrated quantum photonics. Yet, a quantitative theoretical framework connecting realistic waveguide parameters to measurable non-classical photonic output is absent. Here, we present a comprehensive framework for predicting and benchmarking quantum photon statistics in polaritonic integrated circuits, using state-of-the-art experimentally achieved device parameters for (Al)GaAs waveguide platforms. By mapping the pulsed nonlinear waveguide dynamics onto a bosonic quantum circuit representation that explicitly incorporates dissipation, we identify experimentally accessible quantum signatures across two circuit configurations: a single waveguide in a free-space interferometric configuration and a fully integrated multimode coupled-waveguide circuit. We further show that slow-light engineering of the polariton dispersion offers a practical route to amplifying the effective nonlinearity, pushing quantum signatures beyond Gaussian statistics.
\end{abstract}

\maketitle

\section{Introduction}
\label{sec:intro}
The optical Kerr-type nonlinearities reachable under strong coupling between a photonic mode and an excitonic transition~\cite{vladimirova_polariton-polariton_2010} have shown impressive strength in comparison with conventional platforms~\cite{estrecho_direct_2019}. In this strong-coupling regime, the new eigenstates of the system are hybrid light-matter excitations called exciton-polaritons (polaritons). The interplay between their excitonic and their photonic part results in a rich phenomenology and makes them an ideal platform for studying the physics of interacting bosons in optically driven solid-state systems~\cite{carusotto2013quantum}. Polaritonic systems have been used for studying out-of-equilibrium Bose-Einstein condensation~\cite{kasprzak2006bose,deng2010exciton}, supefluidity~\cite{amo2009superfluidity,lerario2017room,ballarini2020directional}, optical parametric generation~\cite{savvidis_angle-resonant_2000-1}, turbulence~\cite{panico_onset_2023} and nontrivial topological phases~\cite{st-jean_lasing_2017, ardizzone_polariton_2022}. Nevertheless, harnessing polariton-polariton interactions in the few particle regimes, a vital ingredient for quantum information processing, has remained quite elusive, despite several theoretical proposals. Recently, few-particle polaritonics has regained experimental interest, especially thanks to the polariton waveguide geometry~\cite{walker_ultra-low-power_2015}. In the polaritonic waveguides, the excitonic transition is strongly coupled to a guided electromagnetic mode confined by total internal reflection (TIR). Although this platform is more sensitive to external perturbations than conventional microcavity polaritons (which need thick Bragg mirrors for building optical cavities to confine the photonic mode), the modular design of waveguide geometries unlocks several new handles for polaritonic processing with strong phase coherence. This includes the on-chip integration of single-photon emitters~\cite{elshaari_hybrid_2020} and squeezing~\cite{park_single-mode_2024}, the design of photonic crystals for dispersion engineering (e.g., slow light~\cite{boyd2009slow}) and applying external electric fields for imprinting a net exciton dipole moment, which adds dipolar interactions to intrinsic polariton nonlinearities~\cite{rosenberg_strongly_2018, suarez-forero_enhancement_2021, liran2024electrically}. These findings have motivated further experimental studies looking for quantum signatures in polaritonic systems. 

Slight effects of antibunched photon statistics of the order of $5\%$ have been reported (just within the experimental error margins) in strongly confined cavity modes~\cite{delteil2019towards,munoz-matutano_emergence_2019} and comparable values have been suggested in waveguide structures by exploiting dipolar interactions ~\cite{ordan_electrically_2024}, thus entering the so-called \emph{polariton-blockade regime}. Consequently, the coexistence of reported strong and tunable nonlinearities and the almost complete absence of clearcut signals related to non-classical photon statistics is the subject of an intense debate. An increasing attention is being directed towards designing experiments in which the quantum regime could be achieved even using smaller nonlinearities and without realizing the full polariton-blockade regime. 

Recently, numerical optimization work has been devoted to the effective mapping of the coupled waveguide field dynamics of laser pulses to a quantum model of coupled nonlinear bosonic modes~\cite{van2025optimizing}. Starting from the procedure presented there, this work aims to use the circuit representation of the quantum photonic integrated circuits (qPIC) to benchmark practical use-cases for establishing non-classical and sub-Poissonian photon statistics in integrated waveguide structures in (Al)GaAs. Strongly attaining to experimentally reported values, we aim to bring the proposed structures within the reach of today's experimental labs, using state-of-the-art fabrication methods as a baseline.

The manuscript is structured as follows. We first illustrate in Sec.~\ref{sec:experimental} the  configuration used as a starting point for the theoretical simulations of the waveguides, based on recently reported experimental values in the literature. Additionally, we summarize the characteristic photonic quantum properties that we aim to discern in experiment for calibrating the non-classicality of the emitted light. In Sec.~\ref{sec:single-waveguide}, we start from the simplest instance in which first non-classical photon statistics can be witnessed, a single waveguide in a free-space interferometric setup. Next, in Sec.~\ref{sec:PIC}, we move to the fully integrated setup. First, the simulation benchmarking of a quantum polaritonic integrated circuit (qPIC) with uniform couplers is addressed, a design planned to be reachable with current platforms, as reported in literature. Finally, we present a pathway for amplifying the non-classicality of the photon statistics in a controllable fashion, by reducing the polaritonic group velocity using techniques of slow light~\cite{baba2008slow,boyd2009slow}, as was recently proposed for amplifying the nonlinearity in integrated GaAs-based structures \cite{rahmani2026strongly}.

\section{Experimental parameters for the waveguide designs}
\label{sec:experimental}
The system considered here is one of the standard platforms for polaritonics, that is, a III-V heterostructure formed by GaAs/Al$_{x}$Ga$_{1-x}$As alloys. More specifically, the waveguide considered here comprises three 7nm-thick GaAs quantum wells (QWs), separated by 3~nm Al$_{0.95}$Ga$_{0.05}$As barriers. The bottom cladding consists of a 1.23um thick Al$_{0.95}$Ga$_{0.05}$As layer. The thicknesses are designed to have the QWs close to the maximum of the TE0 electric field mode.   

\begin{figure}[htbp]
    \centering
    \begin{subfigure}[b]{\linewidth}
        \includegraphics[width=.8\linewidth]{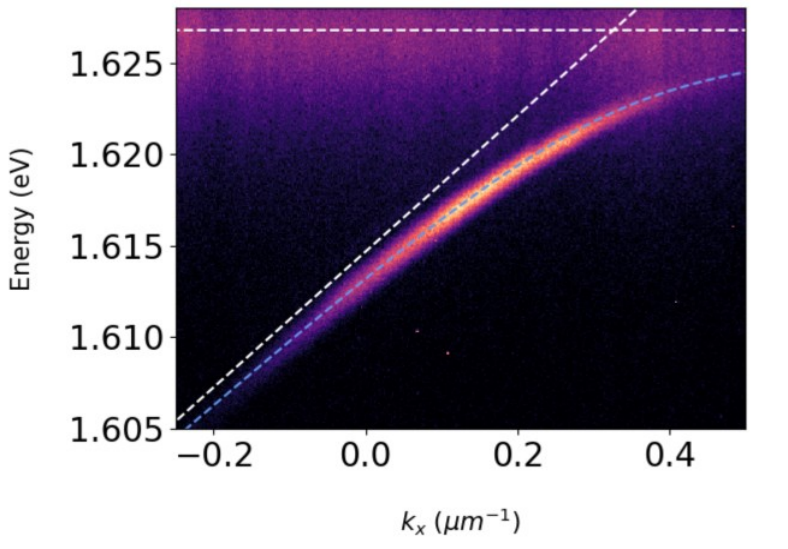}
        \caption{Photoluminescence of LP dispersion}
        \label{fig:disp_exp}
    \end{subfigure}
    \hfill
    \begin{subfigure}[b]{\linewidth}
        \centering
        \includegraphics[width=\linewidth]{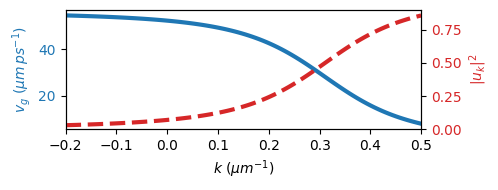}
        \caption{Group velocity $v_g$ and exciton fraction $|u_k|^2$}
        \label{fig:vg_ex_frac}
    \end{subfigure}
    
    \caption{The experimental LP dispersion in an AlGaAs waveguide with GaAs quantum wells and derived quantities. (a) Photoluminscence of the dispersion of the guided polaritonic mode extracted from an input/output diffraction grating; vertical axis shows the energy of the polariton mode while the horizontal axis is the wavevector along the propagation direction. 
    (b) Group velocity $v_g$ (blue solid, left axis) and exciton fraction $|u_k|^2$ (red dashed, right axis) obtained from the dispersion in (a). }
    \label{fig:disp}
\end{figure}

Fig.~\ref{fig:disp}(a) shows the experimental measurement of the dispersion of the lower-polariton mode (LP). The derived LP group velocity and the exciton fraction are displayed in Fig.~\ref{fig:disp}(b). The anticrossing between the linear photonic mode and the excitonic resonance is clearly visible, highlighting the strong light-matter coupling regime, with a Rabi splitting of about $\hbar \Omega = 6$~meV. In the following, we use experimental parameters extracted from this sample as a building block for nonlinear quantum optics devices. 

The integrated waveguides with strong coupling to the excitonic resonance in the semiconductor structure give rise to effective two-body interactions for the lower polariton (LP) modes~\cite{vladimirova_polariton-polariton_2010}, mediating a photonic Kerr-type nonlinearity in the integrated waveguides. 

More precisely, regardless of the exact physical mechanism underlying the photonic nonlinearity, phenomenologically it stems from a (repulsive) two-body point-like exciton-exciton interaction,

\begin{align}
\label{eq:Hexcexc}
\hat{H}_\text{int}(\vec{k}) =& \frac{\hbar g_\text{exc}}{2} \times \nonumber\\
\int_\mathcal{S} d^2\vec{r}\;& \hat{\psi}^\dagger_\text{exc}(\vec{r}) \,\hat{\psi}^\dagger_\text{exc}(\vec{r})\,  \hat{\psi}_\text{exc}(\vec{r})\,  \hat{\psi}_\text{exc}(\vec{r}) ,
\end{align}

\noindent with $\hat{\psi}_\text{exc}(\vec{r})$ the bosonic exciton field operator, defined in the plane $\mathcal{S}$ with $\vec{r}\equiv(x,y)$ the two planar coordinates, and $\hbar g_\text{exc}$ the GaAs exciton interaction constant, by default found around $\hbar g=6-10\,\mu$eV~$\mu$m$^2$ in absence of electric fields~\cite{vladimirova_polariton-polariton_2010, estrecho_direct_2019}. 

The photonic nonlinearity is mediated through the lower-polaritonic (LP) field, which, under strong coupling, can be represented as a superposition of the exciton field $\hat{\psi}_\text{exc}$ and the photon field $\hat{\psi}_\text{ph}$; 
\begin{equation}
    \hat{\psi}_{\text{LP}, \vec{k}} := u_\vec{k} \hat{\psi}_\text{exc} + v_\vec{k} \hat{\psi}_\text{ph},
\end{equation}
with $u_\vec{k}$ and $v_\vec{k}$ the Hopfield coefficients, normalized as $u_\vec{k}^2 + v_\vec{k}^2=1$ and $\vec{k}\equiv(k_x, k_y)$ the 2-D reciprocal vector  -- see Ref.~\cite{carusotto2013quantum} for an overview. 

From here onward, only the 1-D longitudinal wavevector $k:=k_x$ of the waveguide will be considered, assuming the waveguide confined transverse component $k_y$ has been integrated out. The LP field passing through the waveguide experiences a nonlinearity from the exciton-exciton interaction, $\hbar g_\text{LP}(\vec{k})=u_\vec{k}^4\,\hbar g_\text{exc}$ which, under waveguide excitation, has an explicit dependence on the wavevector $k$, where $u_\vec{k}^2$ represents the exciton fraction of the LP field~\cite{walker_exciton_2013} -- see Fig. \ref{fig:disp}(b), red dashed line.

The presence of nonlinearities in waveguide polaritonic systems has been the subject of intense research activity with the demonstration of soliton formation~\cite{walker_ultra-low-power_2015} and frequency generation~\cite{walker_spatiotemporal_2019}. Nevertheless, these studies have been almost exclusively focused in the classical, mean-field regime, when a high polariton intensity is excited in the waveguide, while the emergence of non-classical quantum correlations from polaritonic light-matter coupling, at the few-photon level, remains largely unreported experimentally. 

To reiterate a key point from the introduction: integrated polaritonic waveguides constitute a modular and scalable platform uniquely suited for engineering non-classical quantum statistics of light~\cite{scala_deterministic_2024}. Crucially, the application of an external electric field activates dipolar exciton-exciton interactions and can dramatically amplify the effective nonlinear rate of the photonic signal~\cite{rosenberg_strongly_2018, suarez-forero_enhancement_2021, liran2024electrically}, well beyond the intrinsic material nonlinearities of the semiconductor~\cite{estrecho_direct_2019}. This aspect will be central to the results presented in this work.

In what follows, we attain to experimentally reported values for the exciton-exciton interaction constant in literature. We start from the generally accepted value in unbiased GaAs samples, $\hbar g=10\, \mu$eV $\mu$m$^2$~\cite{estrecho_direct_2019} and extend this to values reported in literature in presence of  electric fields: $\hbar g=50\, \mu$eV $\mu$m$^2$~\cite{suarez-forero_enhancement_2021} and $\hbar g=700\, \mu$eV $\mu$m$^2$~\cite{liran2024electrically}. 

We note that the interaction constants $\hbar g$, as given here, represent the effective measured value reported in the literature, integrated over the full multi-quantum-well structure. In that sense, the upcoming simulation results rather allow for an order-of-magnitude comparison with the values presented. Indeed, the precise magnitude of $\hbar g$ depends sensitively on the number of quantum wells, their spatial arrangement, and the inter-well spacing. Furthermore, to obtain the reported interaction values, small modifications of the waveguide structure are required. For example, thicker QWs are more adapted to observe dipolar interactions.  These features, along with the dispersive properties, must be carefully assessed, before translating the presented results to the specific (Al)GaAs-based sample at hand. At any rate, these adjustments are expected to result in small modification of the polaritonic properties, in the first place for the LP group velocity. Therefore, this will not have large (qualitative) consequences for the main results described in this work.

As a final note, it was observed recently observed by M. Richard et al.~\cite{richard2026excitonic} that the saturation of the light-matter coupling could also contribute to the optical nonlinearity. They reported a close matching with a photonic saturation of the oscillator strength, in a sample with one single QW. In the context of this work, we believe the low-power waveguide excitation in combination with the multi-quantum-well structure will not cause immediate saturation effects. Hence, we expect the exciton-exciton exchange interaction, of the form given in Eq. \eqref{eq:Hexcexc}, to remain the dominant factor contributing to the polaritonic nonlinearity.

\section{Antibunched Photon Statistics using a Single Polariton Waveguide}
\label{sec:single-waveguide}

\begin{figure}
    \centering
    \includegraphics[width=1\linewidth]{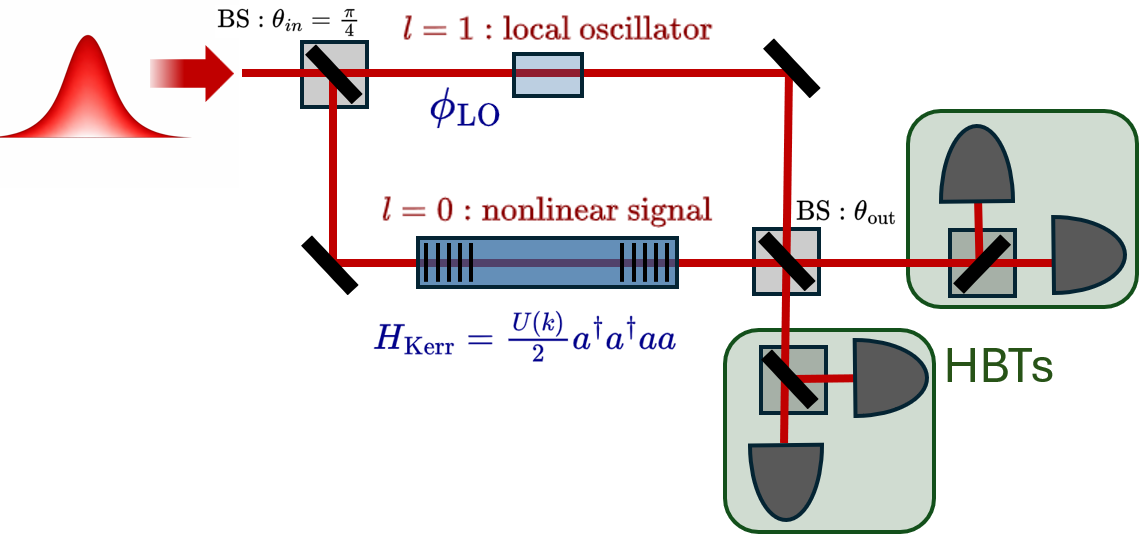}
    \caption{The Mach-Zehnder interferometer setup for generating antibunched photon statistic. After the first $50:50$ beam splitter (BS) coupler, one arm (the local oscillator) has a tunable phase shift $\phi_\text{LO}$, while the other couples in the nonlinear waveguide. After the second coupler a HBT experiment is performed of the signal in the two arms, labeled with the mode index $l=0,1$.}
    \label{fig:single-WG-setup}
\end{figure}

We start with characterizing the possibilities of in- and out-coupling into a single (uncoupled) integrated waveguide device, within a setup of free-space interferometry. The waveguide pulse dynamics is governed by the local nonlinear Hamiltonian, written in the rotating frame at the pump frequency, with $\Delta= \hbar\omega_0(k)-\hbar\omega_\text{pump}$ denoting the energy detuning of the pump w.r.t. the LP waveguide dispersion,
\begin{equation}
\label{eq:Kerr_gate}
    H_\text{Kerr}(k) = -\Delta \,a^\dagger a + \frac{\hbar U(k)}{2} \hat{a}^\dagger \hat{a}^\dagger \hat{a} \hat{a}.
\end{equation}
 Here, the operators $\hat{a}$ and $\hat{a}^\dagger$ represent the bosonic annihilation and creation operator of the photonic field in the pulse. The parameter $U(k)$ is the waveguide pulse nonlinear rate stemming from the LP nonlinearity $\hbar g_\text{LP}(k)$  -- see Appendix \ref{app:parameters} for more information. From here onward, we assume on-resonance excitation of the LP mode so that $\Delta=0$.

Hamiltonian~\eqref{eq:Kerr_gate} is diagonal in photon-number basis and, consequently, does not alter the photon-number output statistics by itself. Instead, it gives nonlinear number-dependent phase shifts $\Delta \phi_n(t)=t\cdot\tfrac{U}{2}n(n-1)$, with $t$ the time of signal coupling into the waveguide. For that reason, there is at least one interferometric operation needed to break the photon-number invariance and redistribute the number states across two separate photonic modes. 

Thus, the first setup that we explore for extracting non-classical statistics from the waveguide output photonic field is a free-space Mach-Zehnder interferometer (MZI) with waveguide coupling in one arm -- see Fig.~\ref{fig:single-WG-setup}). The two signals from the output coupler (beam splitter) are analyzed in a Hanbury Brown-Twiss (HBT) setup for detecting the (symmetric) intensity-intensity correlation matrix, $g^{(2)}_{lm}:=\langle: \hat{n}_l \hat{n}_m :\rangle/ (n_ln_m)$, with $n_l = \langle \hat{n}_l\rangle$ and $\langle: \hat{O}  :\rangle$ indicating normal operator-ordering and $l$ and $m$ labeling the two photonic signals in the MZI node.

The photonic signal of the two output arms, as indicated on Fig.~\ref{fig:single-WG-setup} is represented by the sequence of unitary operations,
\begin{equation}
    |\psi_\text{out}\rangle = U^{\text{(out)}}_\text{BS} \,U_\text{MZI} \,U^{\text{(in)}}_\text{BS}\;|\psi_\text{in}\rangle,
\end{equation}
with $|\psi_\text{in}\rangle=|\alpha_\text{in}\rangle \otimes |0\rangle$ defined as the coherent laser input in the first (symmetric) input arm ($l=0$) of the dielectric coupler, where $\alpha_\text{in}$ represents the amplitude. 

The dielectric free-space couplers are parametrized as the unitary two-mode transform,
\begin{equation}
\label{eq:U_BS}
    U_\text{BS}(\theta) = \exp\Big\{\theta\, \big(\hat{a}^\dagger_0 \hat{a}_1 - \hat{a}_0 \hat{a}^\dagger_1\big) \Big\}\times\exp\big\{i\pi\, \hat{a}^\dagger_1 \hat{a}_1\big\},
\end{equation}
with $\theta$ the mixing angle, and a $\pi$-phase shift in the second, antisymmetric arm (for example, coming from the dielectric coating). We set the in-coupler to $\theta_\text{in}=\frac{\pi}{4}$, i.e., a balanced 50:50 splitter, and the out-coupler to $\theta_\text{out}$ as variable.

The second unitary operation, once the two signals are coupled into the MZI, is defined as,
\begin{equation}
\label{eq:U_MZI}
    U_\text{MZI} = \exp\bigg\{-i\left(\tfrac{U\Delta t}{2} \, \hat{a}^\dagger_0 \hat{a}^\dagger_0 \hat{a}_0 \hat{a}_0 + \phi_\text{LO}\, \hat{a}^\dagger_1 \hat{a}_1\right) \bigg\} .
\end{equation}
Here, $U$ is the waveguide nonlinearity (see Eq. \eqref{eq:Kerr_gate}) in arm $0$ and $\Delta t=\Delta x/v_g$ is the coupling time, influenced by the group velocity $v_g$ and the waveguide length $\Delta x$. The first arm of the MZI ($l=0$) represents the nonlinear signal caused by waveguide coupling, while the second arm ($l=1$) serves as the local oscillator (LO), with a (modular) relative phase $\phi_\text{LO}$, before passing through the out-coupler.

\begin{figure}
    \centering
  
    \includegraphics[width=\columnwidth]{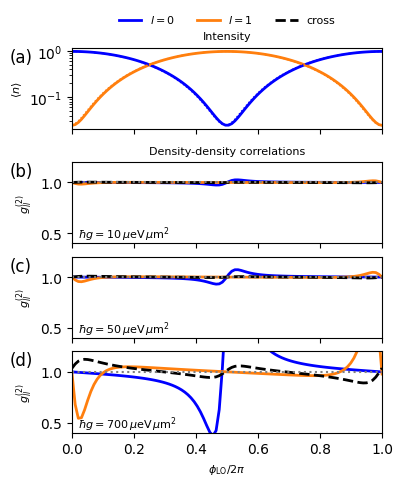}
    \caption{Results for the out-coupled photon statistics of a BS with coupling $\theta_\text{out}=0.2\pi$, when modulating the MZI LO phase within the range $\phi_\text{LO}\in\{0,2\pi\}$ for an incoming pulse of $|\alpha|^2=1$ and $1\,mm$ waveguide coupling at $u_k^2=0.3$ exciton fraction. (a) The varying output intensity across the two modes, on log-scale. (b)-(d) Intensity correlations, intra-mode (blue–orange, solid), cross-correlations (black, dot-dashed) and the coherent correlation for reference (gray, dotted), at the reported nonlinearities. Sub-Poissonian statistics, reaching $g^{(2)}_{ll}<0.5$ at $\hbar g=700\,\mu eV \mu m^2$ is observed in the antisymmetric output arm around $\phi_\text{LO}/2\pi\approx 0.45$, just below $\phi_\text{LO}=\tfrac{\pi}{2}$, after which a sharp turnover to signal bunching occurs.}
    \label{fig:single-WG-results}
\end{figure}

In Fig.~\ref{fig:single-WG-results}, we show the simulated experimental outcomes for $1$ ps pulsed-laser excitation, with $1$ photon on average per pulse ($\alpha_\text{in}=1$). By modulating the optical path length in the second arm, $\phi_\text{LO}$ is scanned in experiment and this alters the redistribution of photons across the two MZI output arms. In Fig.~\ref{fig:single-WG-results}(b)-(d), we consider $U \Delta t = [0.001, 0.005, 0.06]$ for the waveguide nonlinearity, which, as detailed in Appendix~\ref{app:parameters}, correspond to $1000~\mu$m waveguide propagation, with $v_g=40\, \mu m / ps$ and $u_k^2=0.3$ exciton fraction (the value $k=0.23\,\mu m^{-1}$ from Fig.~\ref{fig:vg_ex_frac}), and planar nonlinearity $\hbar g= [10, 50, 700]\, \mu eV \mu m^2$, following the experimental reportings mentioned earlier. In what follows, we present the unitary case for MZI out-coupling of $\theta_\text{out}=0.2\pi$, with no coupling losses.

Varying the LO phase $\phi_\text{LO}$ in experiment causes a redistribution of the photonic output statistics across the MZI output arms. This affects both the measured intensities, $n_l$, and the intensity-intensity correlation matrix, $g^{(2)}_{lm}$. When comparing the signal intensities (Fig.~\ref{fig:single-WG-results}(a)) and the expected density-density correlation (Fig.~\ref{fig:single-WG-results}(b)-(d)), it is clear that strong antibunching (low intra-mode values $g_{ll}^{(2)}$) occurs at weak output intensity, with $g^{(2)}_\text{min}=[0.97, 0.93, 0.37]$ found at $n\approx0.01$ in the first arm ($l=0$, blue line). Additionally, a small nontrivial cross-correlation across the two arms is found (black dot-dashed line). 

In Appendix~\ref{app:MZIloss_coupling}, we discuss in full detail (i) the effect of varying the out-coupling $\theta_\text{out}$ and (ii) the losses of waveguide in- and out-coupling, using the bosonic amplitude-damping Kraus operator representation \cite{leviant2022quantum}, detailed in Appendix~\ref{app:kraus-amplitude-damping}.
 Under strong waveguide-coupling losses, the lower signal strength in one arm gives rise to stronger antibunching effects. However, this comes at the cost of low signal strength. To give an example, with $\hbar g=50\, \mu eV \mu m^2$ and $2.9$ dB coupling losses, values as low as $g^{(2)}_{ll}\approx0.7$ are found, but they come at a signal strength of $\langle n \rangle \approx 10^{-3}$ -- that is, roughly 1 detector click every 1000 pulses. We note that using modern pulsed lasers with a repetition rate of 80 MHz, the signal intensities are still well above the dark counts rate of modern single-photon detectors, so that the intensity correlation measurements are well within experimental reach. 

In addition to implementing the MZI using free-space interferometry, the nonlinear MZI node can also be fully integrated on chip, using symmetric integrated couplers. As a core difference, the (linear) coupling is now represented by the symmetric unitary,
\begin{equation}
    U^\text{sym}_\text{BS}(\theta) = \exp\Big\{ i\theta\, \big(\hat{a}^\dagger_0 \hat{a}_1 + \hat{a}_0 \hat{a}^\dagger_1\big) \Big\},
\end{equation}
rather than the dielectric representation from Eq. \eqref{eq:U_BS}, and the nonlinear coupling, represented by Eq.~\ref{eq:U_MZI}, occurs symmetrically and continuously in both MZI arms during the coupling. On-chip phase shifts $\phi_\text{LO}$ are realized as imprinted optical path-length differences, or by externally applied electrical fields~\cite{rosenberg_strongly_2018,suarez-forero_enhancement_2021, ardizzone2026fewphoton}. 
In Appendix~\ref{app:integratedMZI}, we characterize the the integrated nonlinear MZI node and find qualitatively similar results as expected for the free-space setup from Fig.~\ref{fig:single-WG-results}. However, one advantage of an integrated MZI is the expected stability towards losses, for symmetric in- and out-coupling losses in both arms. Indeed, we explain analytically in Appendix~\ref{eq:g2gauss} how values for $g^{(2)}_{ll}$ are robust towards signal losses, uniform for all modes, in space and time, at sufficiently low nonlinearities $U\Delta t$.

\section{Circuit-generated non-classical photon statistics}
\label{sec:PIC}

In this section,  we present the possibilities of observing non-classical photon statistics in the output of a fully integrated optical circuit, consisting of multiple waveguides coupled into an MZI mesh architecture; a quantum photonic integrated circuit (qPIC).

\subsection{Mapping the field dynamics to a quantum circuit model}

\begin{figure}
    \centering
    \includegraphics[width=0.9\linewidth]{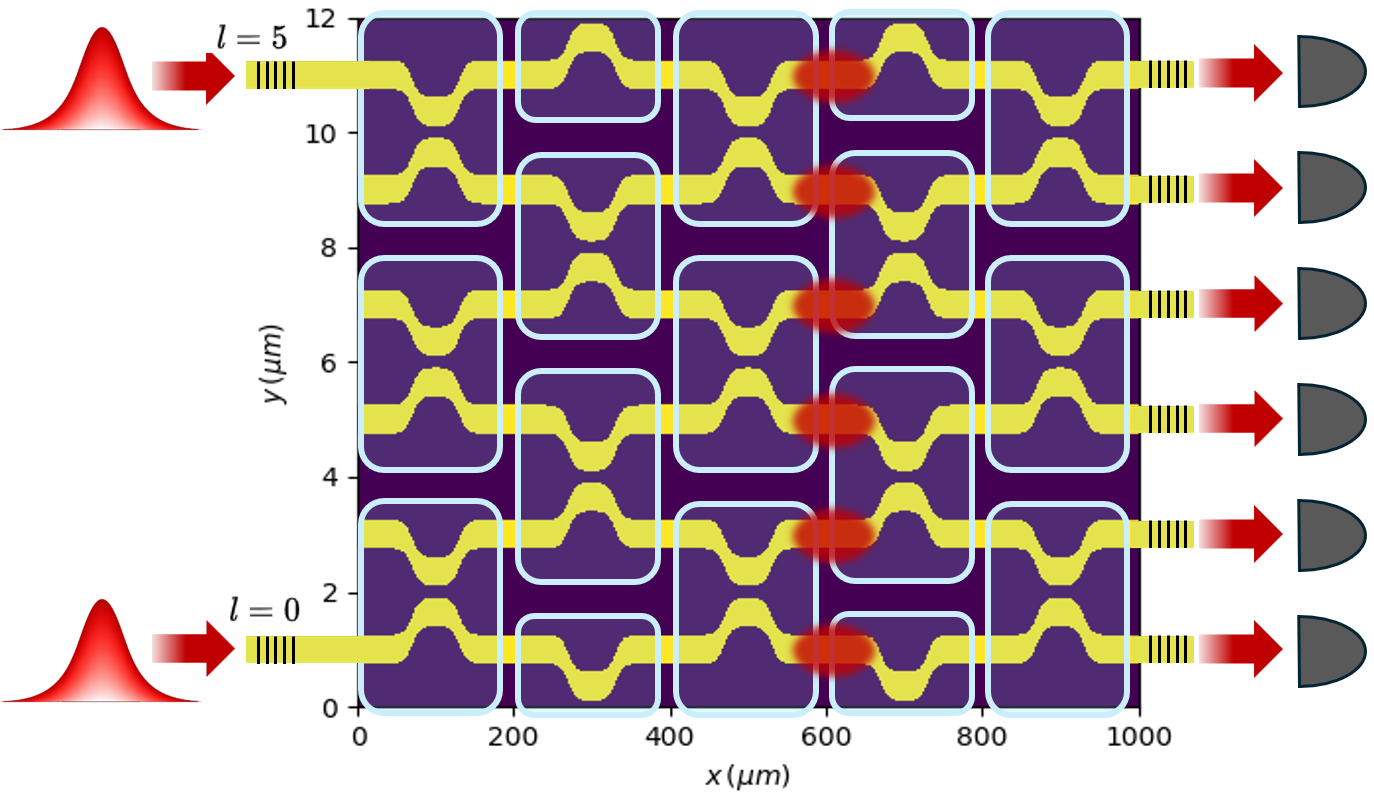}
    \caption{The prototype architecture studied for the first experiments. It consists of $L=6$ coupled waveguides, integrated in a $1\,mm$ architecture consisting of $D=5$ stacked layers of coupling gates. The signal is injected in waveguides $l=0$ and $l=5$ and the output signals of the waveguides are detected individually. It is assumed that all couplers are uniform across the circuit, but the circuit coupling $J\Delta t$ and $U\Delta t$ will rescale in function of the input wavevector $k$.}
    \label{fig:circuit-architecture}
\end{figure}

Under well-defined conditions, the polaritonic pulsed field $\hat{\psi}_\text{LP}(\vec{r})$, propagating in the waveguide with index $l$, is represented as an effective bosonic circuit mode, defined by the bosonic creation and annihilation operators $\hat{a}_l^\dagger$ and $\hat{a}_l$, respectively, inserting or removing a single polaritonic excitation from the waveguide pulse -- see Appendix of~\cite{van2025optimizing}. 

The symmetric integrated couplers are described as two-mode nonlinear coupling gates. Acting on adjacent the modes $l$ and $l+1$, in circuit layer (depth) $d$, the gate is represented by the Hamiltonian,

\begin{equation}
\label{eq:H2}
\begin{aligned}
\hat{H}_{l,d}^{(2)}(k) =\;&
-\Delta \big(\hat{a}^\dagger_l \hat{a}_l + \hat{a}^\dagger_{l+1} \hat{a}_{l+1}\big)\\
&
- \hbar J_{l,d}
\left(
    \hat{a}^\dagger_{l+1} \hat{a}_l
    + \hat{a}_l \hat{a}^\dagger_{l+1}
\right) \\
&
+ \frac{\hbar U(k)}{2}
\left(
    \hat{a}^{\dagger 2}_l \hat{a}^2_l
    + \hat{a}^{\dagger 2}_{l+1} 
      \hat{a}^{2}_{l+1} 
\right).
\end{aligned}
\end{equation}

Here, $\Delta$ is the detuning of the incident laser light from the LP resonance, which, as before, we set to $\Delta = 0$. The temporal coupling rate $J_{l,d}$ represents the symmetric tunneling between waveguides $l$ and $l+1$, as designed by adjusting the waveguide spacing -- see Appendix of Ref.~\cite{van2025optimizing}. The parameter $U(k)$ describes the intra-waveguide nonlinear rate and depends explicitly on the incident wavevector $k$ -- see Appendix~\ref{app:parameters}.

In Fig.~\ref{fig:circuit-architecture}, we show the design considered for the experiment; it consists of $L=6$ coupled waveguides, placed in a $1\,mm$ long circuit architecture, divided in $D=5$ layers of $\Delta x=200\,\mu m$ length each. Correspondingly, the gate times for signal propagation are found as $\Delta t_k=\Delta x/v_g$.

In the simulated experiment, the wavevector $k$ of the photonic signals, uniform across the waveguides $l$, is scanned, in resonance with the LP distribution $\omega(k)$. All circuit couplers are set to uniform, $J_{l,d}\equiv J$, with a calibrated $J\Delta t\equiv\tfrac{\pi}{6}$ in the photonic regime (the limit of low $k$ on Fig.~\ref{fig:disp}, at zero exciton fraction). We consider two in-coupled pulsed signals with amplitude $|\alpha_{0}|=|\alpha_{5}|=1$ and a relative phase difference of $\varphi_\text{rel}=0.6\pi$, in order to break the mirror symmetry $l\leftrightarrow L-1-l$. The pulses have an average of one photon each (Poisson-distributed) and a duration of $1$ ps. 

Both the exciton fraction and LP group velocity vary across the LP branch (see Fig.~\ref{fig:disp}(b)), which rescales the gate couplings $J\Delta t_k$ and the nonlinearity $U(k)\Delta t_k$. Therefore, it is expected that the waveguide output signals have $k$-dependent intensity and intensity auto-correlations, $g^{(2)}_{ll}(k)$. Furthermore, it is important that samples are collected pulse-based, with no cross-pulse time correlations at time delays different from $\tau=0$.

The noisy circuit simulations, including photonic losses, were run using the formalism of the matrix product state density operator (MPDO) representation -- see Ref.~\cite{verstraete2008matrix} for the original work, or Refs.~\cite{muller2024enabling,cheng2021simulating} for details on the noisy circuit simulation. In addition to the unitary gates from Eq. \eqref{eq:H2}, circuit noise in terms of photonic losses can be incorporated using Kraus operators for bosonic amplitude damping~\cite{leviant2022quantum} -- see Appendix~\ref{app:kraus-amplitude-damping} for more details. Before the signal entering the gate, the number of polaritonic loss events is sampled, thereby increasing the MPDO \emph{purity} dimension. The full open-source code for qPIC simulation, implemented in PyTorch~\cite{paszke2019pytorch}, can be found in the GitHub repository~\cite{tensorCircuitTorchGit}. 

In short, we used a local Fock-space truncation at $\mathcal{N}=10$, thus a physical tensor dimension of $d=11$ -- equal to $\mathcal{D}=11^6\approx 10^{6.25}$ full Hilbert-space dimension for the full circuit quantum simulation. The tensor bond and purity dimension both have a cutoff at singular values of $10^{-9}$. The photonic losses before entering a new circuit layer were sampled using the Kraus-operator representation for bosonic amplitude damping -- see Appendix \ref{app:kraus-amplitude-damping}. 

\subsection{Simulated experimental qPIC signal}

\begin{figure}[htbp]
    \centering
   
    \includegraphics[width=\columnwidth]{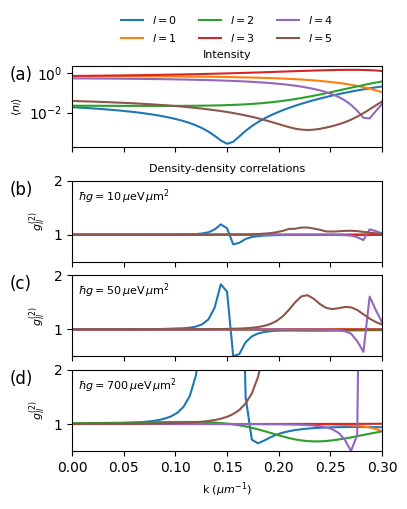}
    
    \caption{The simulation results of a circuit of $L=6$ waveguides and depth $D=5$, for experimentally reported exciton interaction strengths, for zero losses and $\varphi_\text{rel}=0.6\pi$. (a) The output pulse intensities (pulse photon number) for $\hbar g=10\mu eV \mu m^2$ (indistinguishable from $\hbar g=[50,700]\,\mu eV \mu m^2$). (b) The expected intra-pulse intensity-intensity correlations. Strong antibunched photon statistics is witnessed at low output intensities. In all cases, there is sharp crossover from antibunching to bunching statistics around the low-intensity values $k\approx[0.15,\,0.27]\,\mu m^{-1}$}
    \label{fig:g2_k_scan}
\end{figure}


In Fig.~\ref{fig:g2_k_scan}, the results of the circuit simulation are shown, for three values reported in literature for the interaction strength; $\hbar g=10 \,\mu eV\, \mu m^2$ (bare exciton interaction, Ref.~\cite{vladimirova_polariton-polariton_2010}), $\hbar g=50 \,\mu eV\, \mu m^2$ (Ref.~\cite{suarez-forero_enhancement_2021}), $\hbar g=700 \,\mu eV\, \mu m^2$ Ref.~\cite{liran2024electrically}). 

Comparing the three panels, Fig.~\ref{fig:g2_k_scan}(b)-(d), it is clear that both the signal bunching and antibunching are amplified for larger $\hbar g$. In Fig.~\ref{fig:g2_k_scan}, we show the intensity shifts for $\hbar g= 10\, \mu eV \mu m^2$, on log-scale. Intensity shifts for the other values of $\hbar g$ are not distinguishable with the bare eye. Interestingly, there is a sharp crossover around $k_\text{cross} \approx 0.15$ where the signal $l=0$ changes abruptly from strong antibunching to strong bunching. The minimal value retrieved from the figures is $g^{(2)}_\text{min}=0.82$ at $\langle n\rangle = 3.3\cdot10^{-4}$ mode output intensity for $\hbar g=10 \,\mu eV\, \mu m^2$, $g^{(2)}_\text{min}=0.50$, with $\langle n\rangle=3.3\cdot10^{-4}$ for $\hbar g=50 \,\mu eV\, \mu m^2$ and $g^{(2)}_\text{min}=0.50$ at $2.5\cdot10^{-2}$ output intensity for $\hbar g=700 \,\mu eV\, \mu m^2$. For the latter, it is noted that $g^{(2)}_{ll}$ of waveguide $l=4$ at $k\approx0.27$ drops below the minimal value, otherwise found at $k\approx0.17$ in waveguide $l=0$. We conclude that values of strong signal antibunching are generally reached at lower signal strengths. Put in numbers, $\langle n\rangle=3.3\cdot10^{-4}$ intensity would yield about $3.3$ clicks per 10k pulses. Thus, with a 80 MHz pulsed laser, this gives about 25k clicks per second. For a comparison, commercially available fibered Supercondicting Nanowires Single Photon Detectors (SNSPDs) have much lower dark count rates, on the order of tens of Hz, well below the signal level anticipated in this work. 

 Although the results in Fig.~\ref{fig:g2_k_scan} were obtained using semi-exact MPS/MPDO simulations of the unitary circuit, the resulting photon statistics can largely be understood within the framework of Gaussian quantum states, valid for weak nonlinearities. This conclusion is explained in Appendix~\ref{app:gauss} and follows from two numerical observations; (i) the nonlinearity does not redistribute the output signal strength and (ii) the signal strength (or, equivalently, the uniform photonic loss rate) does not affect the intensity auto-correlations.

These results can be understood within the framework of the \emph{unconventional photon blockade}~\cite{liew2010single,bamba2011origin,flayac2017unconventional}, which has been observed in superconducting hardware~\cite{vaneph2018observation}. The underlying physics can be fully understood from the framework of Gaussian quantum states~\cite{lemonde2014antibunching} and schemes have been proposed to witness the effect in 2-D GaAs microcavities using Fourier-space filtering of the emitted light~\cite{van2018engineering}. The semiclassical, Gaussian nature of the antibunching can be exceeded by engineering larger photonic nonlinearities in the material. In what follows, we discuss the case of using techniques of `slow light' waveguide propagation to achieve this \cite{rahmani2026strongly}.

\subsection{Amplifying the circuit non-Gaussianity using slow light}

\begin{figure}[htbp]
    \centering
   
    \includegraphics[width=\columnwidth]{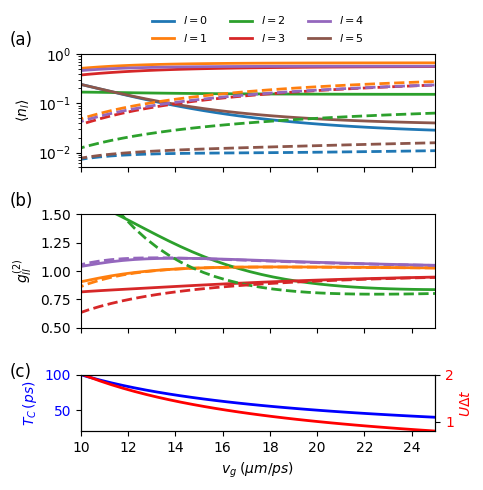}
    
    \caption{The lossless (solid lines) and lossy (dashed lines) simulation results of a circuit of $L=6$ waveguides and depth $D=5$. The LP group velocity $v_g$ is modulated, and a relative pulse phase shift of $\varphi_\text{rel}=0.2\pi$, coupling $J\Delta t = \tfrac{\pi}{6}$, $\hbar g=700\,\mu eV \mu m^2 $ and $u_k^2=0.3$ exciton fraction is considered. For the losses (dashed lines), we consider the (uniform) case of $\gamma=0.02\,ps^{-1}$. (a) The output signal intensities and (b) the values for the intensity auto-correlation. (c) The circuit time of photon propagation (left, blue) and the effective gate nonlinearity $U\Delta t$ (right, red). Low group velocities surpass the Gaussian statistics; output intensities gain a nonlinear dependence and $g^{(2)}_{ll}$ is affected by the losses.}
    \label{fig:g2_vg}
\end{figure}

A promising way to increase the effect of the polaritonic nonlinearity is using techniques of slow light~\cite{yang2011lasing}. Indeed, a lower LP group velocity $v_g$ increases the gate nonlinearity $U\Delta t$ in two manners; slow light evidently leads to longer gate times, $\Delta t = \Delta x/v_g$, but, additionally, also to a spatial compressing of the pulse width, $\sigma_z=\sigma_t/v_g$. In Appendix~\ref{app:parameters}, we explain how this, in first approximation, leads to stronger gate nonlinear rates $U$. However, it must be noted that slow light in photonic structures also has practical limitations. Near the band edge, where the group velocity is minimal, the group-velocity dispersion (GVD) becomes severe, causing temporal broadening and distortion of pulses. Moreover, also the propagation losses will increase with decreasing group velocity, since fabrication disorder and surface roughness scatter slow-moving photons more effectively~\cite{baba2008slow}.

In Fig.~\ref{fig:g2_vg}, the results are shown for the same circuit architecture previously considered ($L=6$ coupled waveguides and $D=5$ circuit depth, $1\,ps$ pulses, but now $\varphi=0.8\pi$ input-pulse phase difference) in which we explicitly slow down the LP group velocity $v_g$ between the limits $10$ and $25\,\mu m /ps$, below the previous lowest value considered, $v_g\approx 20\, \mu m/ps$. The coupling of the circuit gates is set to $J\Delta t=\frac{\pi}{6}$ in the design, uniformly, for any value of $v_g$, and the exciton fraction is $u_k^2=0.3$. In Fig.~\ref{fig:g2_vg}(a), the waveguide output intensities $\langle \hat{n}_l$ clearly display a variation at low $v_g$, before they converge to the values from before at higher $v_g$. This is understood as a sign of a photonic tunneling blockade, or \emph{self-trapping}, under strong nonlinearity~\cite{jensen1982nonlinear}. Indeed, the outer modes $l=0,5$ (blue and brown curve), where the signal is injected, remain equally occupied when $v_g\rightarrow 0$. Fig.~\ref{fig:g2_vg}(b) shows the intensity autocorrelation, which for the modes $l=1,3$ dives below $g^{(2)}_{ll}<1$ at low $v_g$. It will converge back to $g^{(2)}_{ll}=1$ for high $v_g$ (or low nonlinearity), recovering the elementary Gaussian quantum statistics presented earlier. Important, we note that the amount of antibunching (minimal values of $g^{(2)}_{ll}$) are of the same order (or even slightly higher) as found earlier using the experimental LP dispersion from Fig.~\ref{fig:disp}. This not unexpected, since the circuit antibunched output is caused by an intricate balance between the nonlinearity and linear mode interferences. Therefore, optimization schemes for finetuning the individual gate coupling rates can address this issue further (see Ref.~\cite{van2025optimizing}) but this is beyond the scope of this work. Finally, in Fig.~\ref{fig:g2_vg}(c), the resulting circuit time $T_C = D\Delta t$ (left, blue) and gate nonlinearity $U\Delta t$ (right, red) are given for the corresponding group velocity $v_g$. 

From these results, we conclude that using techniques of slow light has a significant impact on the quantum statistics generated by the qPIC, thus opening the way to on-chip quantum manipulation and quantum processing for applications, including quantum-state preparation~\cite{ghosh2019quantum, van2025optimizing}, quantum tomography~\cite{ghosh2020reconstructing} and quantum  sensing~\cite{munoz2024photonic,van2025optimizing}, beyond the limit of Gaussian quantum state processing.

\section{Conclusions}

We presented a comprehensive theoretical framework for benchmarking quantum photonic integrated circuits (qPICs) using polaritonic waveguides, with the explicit goal of identifying experimentally accessible non-classical photon statistics in state-of-the-art (Al)GaAs platforms. After establishing the LP dispersion parameters and the effective bosonic circuit description for the lower-polariton modes, we analyzed two circuit configurations: first, waveguide coupling one free-space MZI arm and, second, the full on-chip circuit integration, using an architecture of stacked symmetric couplers.

First, we demonstrated that the interplay between the Kerr-type waveguide nonlinearity and free-space linear interferometric operations gives rise to a rich $g^{(2)}_{ll}$ profile as a function of the LO phase in the MZI. A sharp crossover from strong antibunching to bunching is found, whose depth scales with the interaction strength $\hbar g$. Second, we analyzed a fully integrated multi-waveguide qPIC with uniform evanescent couplers. By scanning the input wavevector $k$, the LP group velocity and exciton fraction are tuned simultaneously, driving $k$-dependent intensity auto-correlations across the circuit output modes. Crucially, varying the material nonlinear rate amplifies the antibunched and bunched signals, but leaves the output intensities largely invariant. This is similar in nature as the unconventional photon blockade mechanism, operating in the weakly interacting Gaussian regime \cite{flayac_unconventional_2017}. Nevertheless, we highlight an important difference between the work presented here and previous unconventional blockade proposals: in the case considered here, pulsed waveguide excitation, there is no need to perform temporally resolved measurements of the second-order correlation function, which lifts severe constraints on the experimental temporal resolution needed to observe these effects. Finally, we demonstrated that slow-light engineering of the polariton dispersion constitutes a practical route to pushing the circuit response beyond Gaussian statistics: reducing $v_g$ simultaneously extends the photonic gate exposure time and spatially compresses the pulse, thus strongly amplifying the accumulated gate nonlinearities $U\Delta t$.

We stress that the parameter regimes explored throughout this work are well within reach of (or very close to) current waveguide polariton samples, and the proposed experiments can be implemented in the near future.

Moreover, several natural extensions of this work suggest themselves. A direct next step is the experimental validation of the predicted antibunching profiles and their $k$-dependence in existing (Al)GaAs waveguide samples. Starting from our results, small sample-specific modifications will provide concrete and quantitative benchmarks for experiments. On the theoretical side, a systematic optimization of the individual circuit couplers, potentially with integrated on-chip modular phase shifters, is expected to significantly amplify the non-classical sub-Poissonian output. The MPDO framework developed here is directly compatible with gradient-based design tools for this purpose -- see Ref.~\cite{van2025optimizing}. The theoretical modeling, calibration and inclusion of dipolar polariton interactions through applied electric fields~\cite{rosenberg_strongly_2018, suarez-forero_enhancement_2021, ordan_electrically_2024} must be carefully addressed for future experiments: this represents the most immediate pathway to accessing stronger photonic nonlinearities. Looking further ahead, the bosonic circuit description used here is naturally suited for extensions towards multi-photon entanglement generation, quantum state preparation and tomography~\cite{ghosh2019quantum, ghosh2020reconstructing}, and variational quantum sensing protocols~\cite{munoz2024photonic}. This positions polaritonic integrated circuits as a compelling solid-state platform for near-term quantum photonic applications, using deterministic nonlinear integrated circuit designs.

\section*{Acknowledgements}
The authors are grateful to the consortium of the Q-ONE project for collaborative and insightful discussions. V.A. and D.S., A.R. and E.M., M.V.R. and T.V.V. acknowledge support from the following project “Quantum Optical Networks based on Exciton-polaritons”, (Q-ONE, N. 101115575, HORIZON-EIC-2022PATHFINDER CHALLENGES EU project). V.A. and D.S. acknowledge support from “Integrated Infrastructure Initiative in Photonic and Quantum Sciences” (I-PHOQS, N. IR0000016, PNRR MUR project). A.R. and E.M. acknowledge Ailton Garcia Jr., Melina Peter, and Naser Tajik for the growth and characterization of the GaAs/AlGaAs QWs used for Fig.~1 and the support of the Austrian Science Fund (FWF) [10.55776/COE1, 10.55776/PIN4389523, 10.55776/FG5].

\bibliographystyle{unsrt} 
\bibliography{refs} 

\clearpage
\appendix
\section{Derivation of the circuit gate parameters}
\label{app:parameters}

The tensors corresponding to the circuit gates are constructed from the two-mode Hamiltonian, defined on resonance with the LP mode. It includes the symmetric mode coupling rate $J$ and the nonlinearity $U$ (see Eq. \eqref{eq:H2} in main text), which the photonic pulsed field is exposed to for a time $\Delta t = \Delta L/v_g$, with $\Delta L$ the gate length and $v_g$ the LP group velocity, when passing the gate.

The linear gate coupling rate scales as $J\Delta t$, while the gate nonlinearity computed from the pulse envelope and the intrinsic GaAs material 2-D nonlinearity $g=u_k^4\, g_\text{exc}$, with $g_\text{exc}$ the exciton-exciton interaction constant. For this, we find,

\begin{equation}
\label{eq:Uk}
    U(k) = g_\text{LP} \int d\vec{r}\, n^2(\vec{r};k),
\end{equation}
with $n(\vec{r},k)$ the normalized pulse field profile,
\begin{eqnarray}
    n(z,y;k) = \frac{1}{\sqrt{2\pi \sigma_z^2}} e^\frac{-(z-z(t))^2}{2\sigma_z^2(k)}\times \mathcal{F}(y).
\end{eqnarray}
Here, the spatial pulse propagates at $z(t)=v_gt$ and has a width that scales with the group velocity, $\sigma_z(k) = v_g(k)\,\sigma_t$ where $\sigma_t$ is the pulse duration (which we set, by default, to $\sigma_t=1\,\text{ps}$ in the main text). The function $\mathcal{F}(y)$ represents the transversal waveguide confinement of the field. determined by the Helmholtz equation -- see, e.g., Ref.~\cite{snyder1983optical}.

An important note, we assume here that (i) the quantum field contained in the pulse is coherent and (ii) that the pulse envelope remains invariant over time. We are currently investigating in more detail how dispersive and nonlinear effects impact these assumptions. First insights confirm that the pulse envelope can slightly deform under dispersive or nonlinear effects, but the contained field remains coherent. Therefore, these corrections can be accounted for by rescaling the gate nonlinearity $U(k;t)\,\Delta t_k$ over time.

\section{The Gaussian statistics}
\label{app:gauss}
The photon statistics in the weak interaction limit can be largely understood from a Gaussian quantum-state analysis of the pulsed photon field. For this, we approximate the full boson operator of waveguide $l$ as a classical field, imprinted by the coherent laser beam, plus small fluctuations induced by the nonlinear LP coupling, $\hat{a}_l \rightarrow \alpha_l +  \hat{\chi}_l$. Here it is assumed that the fluctuations obey Gaussian quantum statistics.

Expanding the mean photon number and the fluctuations up to quadratic order in the field operators gives,
\begin{eqnarray}
   \langle \hat{a}^\dagger_l \hat{a}_l \rangle &=& |\alpha|^2_l + \delta n_l,\\
   \langle \hat{a}^\dagger_l \hat{a}^\dagger_l \hat{a}_l \hat{a}_l \rangle &=& |\alpha_l|^4 + 4|\alpha_l|^2\delta n_l + 2 \text{Re}\big\{ \alpha^{\ast2}_l \delta c_l\big\}\nonumber,
\end{eqnarray}
with $\delta n_l:=\langle \hat{\chi}^\dagger_l \hat{\chi}_l\rangle$, the density of fluctuations, and $\delta c_l:= \langle \hat{\chi}_l \hat{\chi}_l\rangle$. For the in-coupled coherent field, only the classical mean-field amplitude $\alpha_l$ is occupied, while the fluctuations $\hat{\chi}_l$ build up under the nonlinearity.

The intra-mode density-density fluctuations are found as,
\begin{eqnarray}
\label{eq:g2gauss}
    g^{(2)}_{ll} &=& \frac{|\alpha_l|^2 + 4\delta n_l + 2\text{Re}\{e^{2i\varphi_l}\delta c_l\}}{|\alpha_l|^2 + 2\delta n_l}\nonumber\\
    &\approx& 1 + \frac{2}{|\alpha_l|^2}\text{Re}\big[\delta n_l + \delta c_l e^{2i\varphi_l}]
\end{eqnarray}
with $\varphi_l$ the relative phase between the classical field $\alpha_l$ and the fluctuation $\hat{\chi}_l$.

Next, we study the rate of growth of the fluctuations $\delta n$ under the single-mode quartic interaction Hamiltonian. Up to quadratic order in the fluctuations, we find,
\begin{eqnarray}
    \label{eq:Hquad}
    H_\text{quad} &=& \frac{U}{2} \big(\alpha^\ast + \hat{\chi}^\dagger\big)^2 \big(\alpha + \hat{\chi}\big)^2\nonumber\\
    &\approx& \frac{U}{2} |\alpha|^4 + 2U|\alpha|^2 \chi^\dagger \chi \nonumber\\
    &&+ \frac{U}{2} \big(\alpha^2 \hat{\chi}^{\dagger2} + \alpha^{\ast2} \hat{\chi}^2\big).
\end{eqnarray}
Using Heisenberg equations of motion for the operator $\hat{\chi}$ and taking expectation values for the cumulants we find a set of coupled cumulant equations of motion,
\begin{eqnarray}
    \label{eq:eomgauss}
    i \partial_t \delta n &=& U|\alpha|^2\big(\delta c^\ast - \delta c),\\
    i \partial_t \delta c &=& 4U|\alpha|^2 \delta c + U\alpha^2(2\delta n +1).
\end{eqnarray}
This leads to the linear matrix equations for the cumulant dynamics (choosing $\alpha$ real),
\begin{equation}
    \partial_t \vec{d} = U\alpha^2 \big(L\vec{d} + \vec{b}\big),
\end{equation}
where we defined $\vec{d}:=\big[\delta n, \delta c, \delta c^\ast\big]^T$ to find the matrix,
\begin{equation}
    L=\begin{pmatrix}
        0& i& -i\\
        -2i& -4i& 0\\
        2i& 0& 4i
    \end{pmatrix}.
\end{equation}
The matrix $L$ governs the \emph{Bogoliubov dynamics} of the fluctuations, with imaginary eigenvalues $(0, \pm 2\sqrt{3}i)$, inducing purely oscillatory evolution at radial frequency $\omega_{\pm} = \pm2\sqrt{3} U |\alpha|^2$. For typical parameters $U=0.02$ and $|\alpha|^2 \approx 1$, this yields $|\omega_{\pm}| \approx 0.07\,ps^{-1}$, corresponding to a period $T \approx 100 \,ps$ -- much longer than typical circuit timescales; $T_C \lesssim 20 \;ps$. Consequently, the oscillatory contributions to $\delta n(t)$ and $\delta c(t)$ remain subleading compared to the linear growth driven by $\vec{b} = [0, -1, 1]^T$, yielding $\delta c(t)= -iU\alpha^2 t$ and $\delta n(t)=\big(U\alpha^2\big)^2 t^2$. 

Consequently, from Eq. \eqref{eq:g2gauss}, we find that,
\begin{equation}
    g^{(2)}_{ll}(t) \approx 1 + Ut\,\cos{\big(2\varphi_l - \tfrac{\pi}{2}\big)}.
\end{equation}
Therefore, we see that for short enough circuits with a low nonlinearity $U$, the Gaussian state assumption remains valid and intensity auto-correlations are independent of the classical field intensity $|\alpha_l|^2$, and solely depend on the phase $\varphi_l$ relative to the fluctuation cumulant $\delta c$. This was observed in Fig. \ref{fig:g2_k_scan} from the main text, when sampling the photonic losses.

\section{Kraus operators for bosonic amplitude damping in the qPIC}
\label{app:kraus-amplitude-damping}
We apply the bosonic pure-loss channel (photon-loss channel) acting on a single photonic mode to simulate the polaritonic loss effects in the integrated waveguides. 

We explain the mechanism for a single bosonic mode, following Ref.~\cite{leviant2022quantum}. The annihilation operator $\hat a$ and number operator $\hat n=\hat a^\dagger \hat a$ admits the Kraus decomposition
\begin{equation}
\label{eq:kraus_damping}
\mathcal{N}_L[\kappa](\rho)=\sum_{\ell=0}^\infty \hat L_\ell \rho \hat L_\ell^\dagger,
\qquad
\hat L_\ell=\sqrt{\frac{\kappa^\ell}{\ell!}}\,(1-\kappa)^{\hat n/2}\hat a^\ell,
\end{equation}
where $\kappa = \gamma\Delta t\in[0,1]$ is the loss probability during the time of gate propagation $\Delta t$, with $\gamma$ the polaritonic loss rate (or, equivalently, $\kappa=1-\eta$ with transmissivity $\eta$). 

In general, the series of `clicks' $\ell$ in the time interval $\Delta t$ in Eq. \eqref{eq:kraus_damping} continues up to infinity or, at least, up to the local truncated Fock space defined for the bosonic mode ($\mathcal{N}=11$ considered in the simulations). When setting $\ell_\text{max}=1$ (that is, `click' or `no click') the jump operator formalism of the Markovian quantum trajectory picture is recovered. In our case, by default, we truncate the number of clicks to sample at $\ell_\text{max}=2$, so at most two photonic loss events, per waveguide mode $l$, in the time interval $\Delta t$, which seems a fair value considering the gate times $\Delta t$, the unitary coupling $J$ and nonlinearity $U$, and the (Poissonian) intensity of order $|\alpha_l|^2\approx1$.

\section{Free-space MZI waveguide coupling losses and output beam-splitter mixing angle}
\label{app:MZIloss_coupling}

\begin{figure}
    \centering
    \begin{subfigure}{\columnwidth}
        \centering
        \includegraphics[width=\columnwidth]{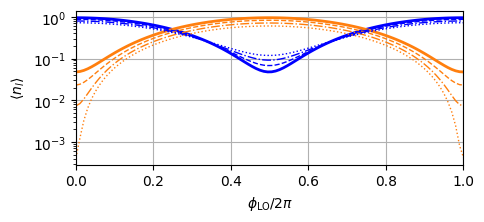}
        \caption{Signal intensity}
        \label{fig:MZI-loss-n}
    \end{subfigure}
    
    \vspace{0.5em}
    
    \begin{subfigure}{\columnwidth}
        \centering
        \includegraphics[width=\columnwidth]{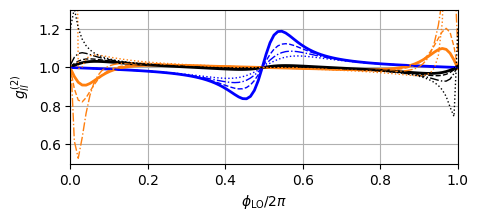}
        \caption{Density autocorrelations}
        \label{fig:MZI-loss-g2}
    \end{subfigure}
    
    \caption{The signal intensity (a) and density autocorrelation (b) in the symmetric (blue lines) and antisymmetric (orange lines) and cross-correlated (black lines) MZI output, considering  $0$ dB (solid), $0.97$ dB (dashed), $1.93$ (dot-dashed) and $2.90$ dB (dotted) coupling losses. Same interferometer as in Fig.~\ref{fig:single-WG-results} of main text, with $U\Delta t = 0.02$ (equivalent to $\hbar g=50\, \mu eV \mu m^2$ in $1\,mm$ waveguide). The signal antibunching is largely amplified under strong losses, due to the lower signal strength in the corresponding MZI arm.}
    \label{fig:MZI-loss}
\end{figure}

We first study in more detail the effects of photonic losses for the in- and out-coupling in the waveguide in one MZI arm, for which the results are shown in Fig.~\ref{fig:MZI-loss}, for the same MZI interferometric setup as described in Fig.~\ref{fig:single-WG-results} in the main text. With large losses, the antibunched photon statistics of the anitsymmetric MZI output arm first gets largely amplified at $\phi_\text{LO}\approx 0$, as a consequence of the weakness of the output signal -- $g^{(2)}_\text{min}\approx 0.55$ at $1.93$ dB loss, at lower signal strength $\langle n_l\rangle\approx 10^{-3}$ (orange dot-dashed line) -- before it shoots up to strong bunching behavior -- see orange dotted line, at $2.90$ dB loss. Thus, we conclude that the low signal strength in one MZI arm first leads to stronger suppression of the 2-photon contribution in the MZI out-coupler, leading to stronger antibunching at a lower photon count rate, before causing severe bunching statistics. On the contrary, around $\phi_\text{LO}/2\pi=0.5$, the effect of the losses is to bring the states closer to coherent states, and still keeping a higher photon flux -- see blue lines. In this case the losses result in a flattening of the $g^{(2)}_{ll}$ autocorrelation function. 

\begin{figure}
    \centering
    \begin{subfigure}{0.48\columnwidth}
        \centering
        \includegraphics[width=\columnwidth]{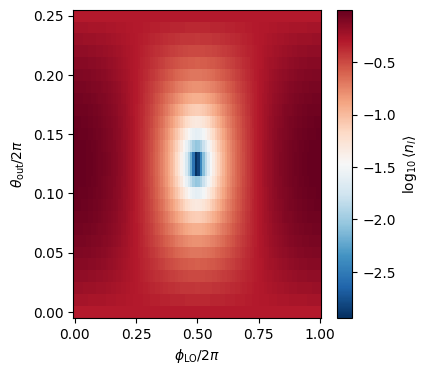}
        \caption{Sym. intensity}
        \label{fig:MZI-loss-n}
    \end{subfigure}
    \hfill
    \begin{subfigure}{0.48\columnwidth}
        \centering
        \includegraphics[width=\columnwidth]{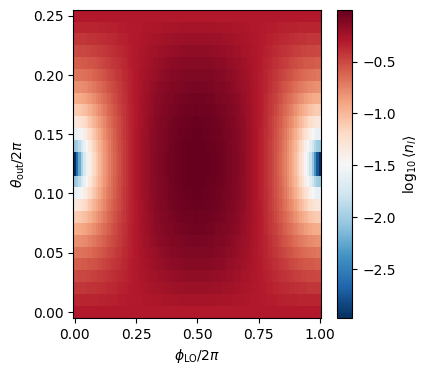}
        \caption{\small Anti-sym. intensity}
        \label{fig:MZI-loss-g2}
    \end{subfigure}

    \vfill
    
    \begin{subfigure}{0.48\columnwidth}
        \centering
        \includegraphics[width=\columnwidth, trim={0cm 0cm 0cm 0cm}, clip]{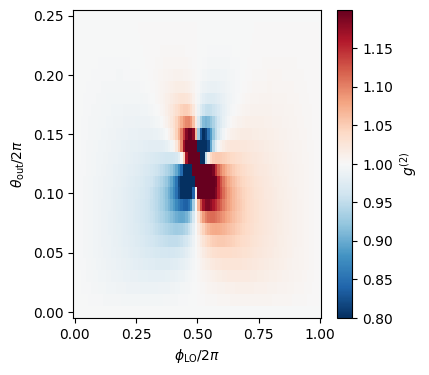}
        \caption{Sym. $g^{(2)}$}
        \label{fig:MZI-loss-n}
    \end{subfigure}
    \hfill
    \begin{subfigure}{0.48\columnwidth}
        \centering
        \includegraphics[width=\columnwidth]{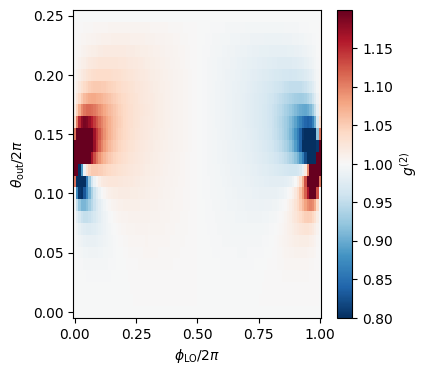}
        \caption{Anti-sym. $g^{(2)}$}
        \label{fig:MZI-loss-g2}
    \end{subfigure}
    
    \caption{The intensity (panel (a)-(b)) and density autocorrelations (panel (c)-(d)) in the symmetric and antisymmetric MZI output arm, for varying local-oscillator phase $\phi_\text{LO}$ and dielectric beam splitter mixing angle $\theta_\text{out}$.}
    \label{fig:theta_phi_MZI}
\end{figure}

Next, we consider the influence of the mixing angle of the second beam splitter, which we previously set to $\theta_\text{out}=0.6\pi$, chosen for convenience, because it gave clear results for showcasing the potential of an MZI with one waveguide element -- see Fig.~\ref{fig:single-WG-results} (main text) and Fig.~\ref{fig:MZI-loss} (above). However, varying the mixing angle (and, hence, the BS splitting ratio) is an important reconfigurable parameter in experiment to consider. For that reason, we present a full 2-D map in Fig.~\ref{fig:theta_phi_MZI} of the density autocorrelation of the symmetric and antisymmetric output arms of the MZI output dielectric beam splitter. Here, the unitary case is presented (no photonic losses) and $U\Delta t=0.02$. It is clear that the photon bunching and antibunching statistics shifts from one output to the other by modulating the parameters $\theta_\text{out}$ and $\phi_\text{LO}$ in experiment. Therefore, we anticipate that it is of crucial importance that they are chosen well, in order to establish a clear antibunched output signal in one MZI output arm, with sufficient signal strength.

\section{An integrated nonlinear MZI node}
\label{app:integratedMZI}

\begin{figure}
    \centering
    \includegraphics[width=\columnwidth]{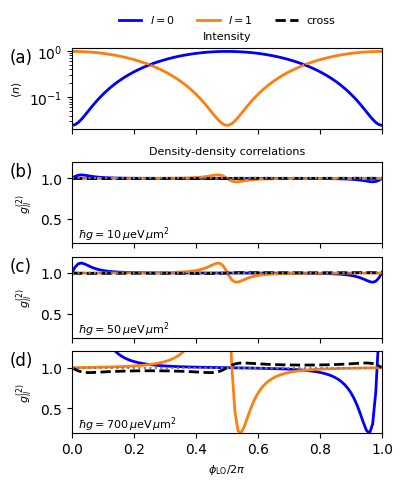}
    \caption{The signal output statistics for a fully integrated MZI. (a) the output intensities, (b)-(d) the intensity correlation matrix elements for $[10, 50, 700]\,\mu eV \mu m^2$, the values from the main text. A sharp crossover from bunching to antibunching is seen around $\phi_\text{LO}$ being multiples of $\pi$. }
    \label{fig:integrated-MZI}
\end{figure}

In the main text, we highlighted that in the integrated waveguide setup a combination of a local (intra-waveguide) photonic nonlinearity must be accompanied be at least one interference to redistribute the photonic particle number across several modes. In Sec.~\ref{sec:single-waveguide}, we illustrated this for waveguide coupling in one arm of a free-space Mach-Zehnder interferometer (MZI), using dielectric beam splitters. However, the MZI setup can also be considered as a fully integrated node, with nonlinear couplers -- that is, one MZI node of the coupled setup later described in Sec.~\ref{sec:PIC}, represented by the Hamiltonian \eqref{eq:H2}. 

In Fig.~\ref{fig:integrated-MZI}, we show the results of the integrated MZI simulation, along the same lines as what was presented in the main text for the free-space setup (Fig. \ref{fig:single-WG-results}). Qualitatively, similar features are shown when scanning the relative phase between the two waveguide signals; $\phi_\text{LO}$. A few remarkable differences, (i) the fully integrated MZI node has a symmetry $\phi_\text{LO}\rightarrow \pi - \phi_\text{LO}$, along with swapping the mode indices $l$; $0 \leftrightarrow 1$ and (ii) strong sub-Poissonian statistics can be reached by finetuning the relative phase $\phi_\text{LO}$ between the waveguides, that is, the optical path length, down to $g^{(2)}_{ll}\approx 0.3$. We note that there is a sharp crossover from strong bunching to strong antibunching around values $\phi_\text{LO}=n\pi$, multiples of $\pi$. The phase shifts can be stabilized in a modular fashion, using the confined Stark effect of the exciton resonance \cite{ardizzone2026fewphoton}. 

Currently, we are investigating in full detail how the two-mode setup can be exploited for future applications in integrated photonic hardware. Loosely speaking, the integrated design has the benefit of facilitating the in-sync processing of the two coupled signals in the MZI node. Using free-space linear interferometry, signal phase alignment (pre and post waveguide coupling) and spectral broadening might cause difficulties for collecting the final correlation measurements.

\section{Influence of the relative phase of the pulses excitation of the qPIC}

\begin{figure*}
    \centering
    \begin{subfigure}[t]{0.32\textwidth}
        \centering
        \includegraphics[width=\linewidth]{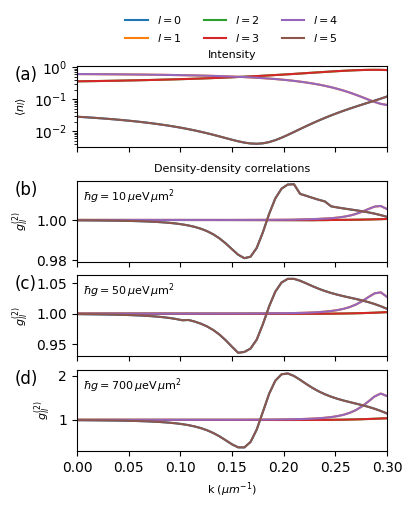}
        \caption{$\varphi_\text{rel}=0.0\pi$}
    \end{subfigure}
    \hfill
    \begin{subfigure}[t]{0.32\textwidth}
        \centering
        \includegraphics[width=\linewidth]{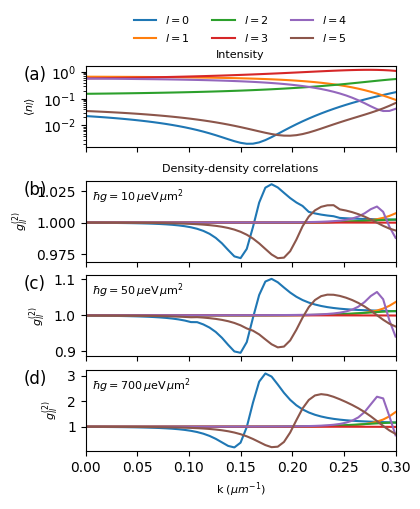}
        \caption{$\varphi_\text{rel}=0.2\pi$}
    \end{subfigure}
    \hfill
    \begin{subfigure}[t]{0.32\textwidth}
        \centering
        \includegraphics[width=\linewidth]{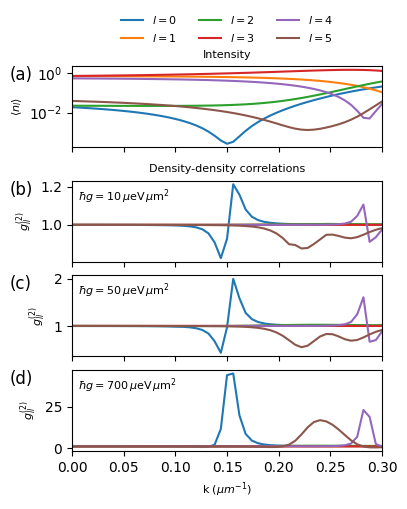}
        \caption{$\varphi_\text{rel}=0.4\pi$}
    \end{subfigure}

    \vspace{0.5em}

    \begin{subfigure}[t]{0.32\textwidth}
        \centering
        \includegraphics[width=\linewidth]{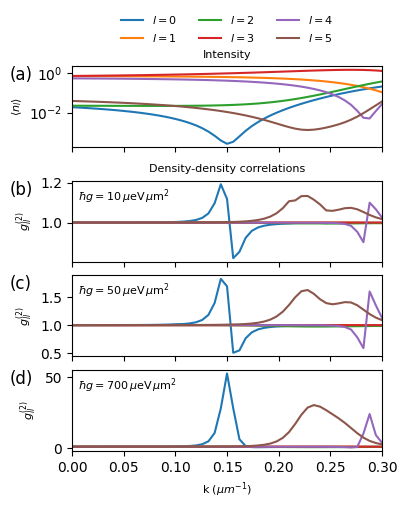}
        \caption{$\varphi_\text{rel}=0.6\pi$}
    \end{subfigure}
    \hfill
    \begin{subfigure}[t]{0.32\textwidth}
        \centering
        \includegraphics[width=\linewidth]{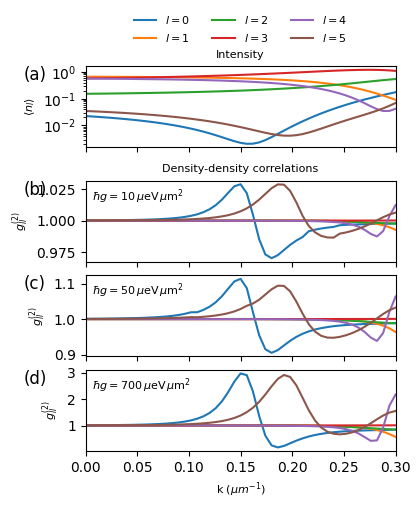}
        \caption{$\varphi_\text{rel}=0.8\pi$}
    \end{subfigure}
    \hfill
    \begin{subfigure}[t]{0.32\textwidth}
        \centering
        \includegraphics[width=\linewidth]{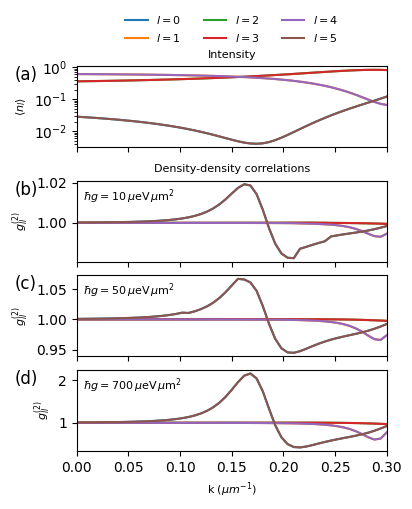}
        \caption{$\varphi_\text{rel}=1.0\pi$}
    \end{subfigure}

    \caption{Results for different relative phase shifts $\varphi_\text{rel}$, for the output intensities (inner upper panels, labeled (a)) and the intensity auto-correlations (inner lower panels, labeled (b)-(c)-(d)) for the experimentally reported interaction constants. Exact multiples of $\varphi_\text{rel}=\pi$ (panel (a) and (f)) result in a platform with mirror symmetry $l\leftrightarrow L-l+1$ so that only $3$ instead of $6$ lines are visible. The case $\varphi=0.6\pi$ (panel (d)) was shown in the main text. Note that the scale of $g^{(2)}$ (y-axis) is given in full here (no cutoff for strong bunching, as in Fig. \ref{fig:g2_k_scan} from the main text)}
    \label{fig:MZI_var_phi_rel_ks}
\end{figure*}

\begin{figure*}
    \centering
    \includegraphics[width=0.8\linewidth]{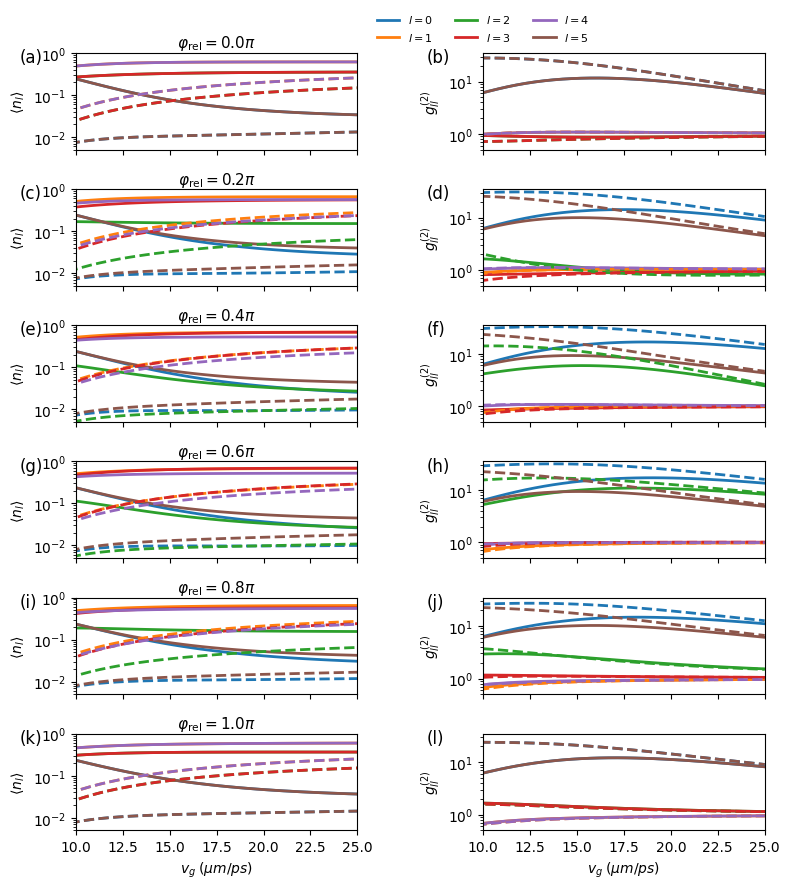}
    \caption{The results for different relative phase shift $\varphi_\text{rel}$ when reducing the pulse group velocity $v_g$. left: intensities, right: intensity auto-correlations, both on log-scale. Full lines: lossless (unitary) circuit, dashed lines: lossy circuit with $\gamma=0.02$ ps$^{-1}$. In the main text, Fig. \ref{fig:g2_vg}, the data from $\varphi_\text{rel}=0.2\pi$ are shown. Note that in that case the strongly bunched signal lines were cut off, in order to highlight the antibunched signal ($g^{(2)}_{ll}<1$). Here we show the full intensity correlations on log scale. }
    \label{fig:MZI_var_phi_rel_vg}
\end{figure*}

In the main text, for clarity, one case of the relative phase shift $\varphi_\text{rel}$ between the two incoming pulses was illustrated for discussing the results of MZI wavenumber scanning (see Fig. \ref{fig:g2_k_scan}) and of the reduced polariton group velocity  (see Fig. \ref{fig:g2_vg}). For the sake of completeness, we provide the full dataset here, with all the relative phases that were simulated. 

In Fig. \ref{fig:MZI_var_phi_rel_ks}, we show the results for different relative pulse phases $\varphi_\text{rel}$ when scanning the wavenumber $k$ in resonance with the LP distribution. In contrast with the main text (Fig. \ref{fig:g2_k_scan}), we did not truncate $g^{(2)}$-values (y-axis) for large values. The low field intensities can give rise to strong bunching behavior, which is particularly visible in panel (d) -- the case given in the main text, but there the data were truncated. Values of $\varphi_\text{rel}$ being exact multiples of $\pi$ (panels (a) and (f)) show a mirror symmetry in density-related outputs (the six lines coincide per two, so only three lines are visible) . This is a consequence of both the circuit and the photon input being perfectly symmetric under the transform $l\leftrightarrow L-l+1$, up to a global phase.

In Fig. \ref{fig:MZI_var_phi_rel_vg}, we present similar data, but now when varying the reduced LP group velocity $v_g$. The intensities are shown on the left and the intensity auto-correlations on the right. For each panel, the lossless unitary case was considered (solid lines) and the lossy case with $\gamma=0.02$ ps$^{-1}$. In the main text, the data from $\varphi_\text{rel}=0.2\pi$ were presented (panels (c) and (d)). For completeness, we provide here the $g^{(2)}_{ll}$ values for the different waveguides on log-scale, to illustrate the strong \emph{bunching} occurrences in certain waveguides, up to $g^{(2)}_{ll}\approx 40$. Qualitatively, the data on for all phase shifts display similar behavior, certain waveguides shift to strong antibunching and others to strong bunching behavior when the group velocity is reduced and, correspondingly, the photonic nonlinearity is increased. Remember, here we considered a rather `brute-force' theoretical approach for probing the accumulated photon statistics given a calibrated integrated circuit architecture. In the future, we plan to work out optimization schemes of the circuit architecture itself, but this is out of scope for this work.

\end{document}